\documentclass[amsmath,amssymb,amsbsy,reprint,pra,preprintnumbers,showpacs,superscriptaddress]{revtex4-1}
\usepackage{graphicx,color}
\usepackage{dcolumn}
\usepackage{bm}
\usepackage{braket}
\usepackage{mathtools}
\usepackage{ulem}
\usepackage[breaklinks,colorlinks=true,linkcolor=blue,urlcolor=blue,citecolor=blue]{hyperref}
\usepackage{times}
\usepackage{physics}
\usepackage{latexsym}
\usepackage{amsmath, amssymb}
\usepackage{mathtools}
\usepackage{multirow}

\newcommand{\be}{\begin{eqnarray}}
\newcommand{\ee}{\end{eqnarray}}

\begin{document}

\title{Bulk-Measurement-Induced Boundary Phase Transition in Toric Code and Gauge-Higgs Model}
\date{\today}
\author{Yoshihito Kuno} 
\thanks{These authors equally contributed}
\affiliation{Graduate School of Engineering Science, Akita University, Akita 010-8502, Japan}
\author{Takahiro Orito}
\thanks{These authors equally contributed}
\affiliation{Insitute for Solid State Physics, The University of Tokyo, Kashiwa, Chiba, 277-8581, Japan}
\author{Ikuo Ichinose} 
\affiliation{Department of Applied Physics, Nagoya Institute of Technology, Nagoya, 466-8555, Japan}


\begin{abstract}
Study of boundary phase transition in toric code under cylinder geometry via bulk projective measurement is reported. 
As the frequency of local measurement for bulk qubits is increased, spin-glass type long-range order on the boundaries emerges indicating spontaneous-symmetry breaking (SSB) of $Z_2$ symmetry.  
From the lattice-gauge-theory viewpoint, this SSB is a signal of a transition to Higgs phase with 
symmetry protected topological order.  
We numerically elucidate the properties of this phase transition in detail, especially its criticality, and give a physical picture using non-local gauge-invariant symmetry operators. 
Phase transition in the bulk is also studied and its relationship to the boundary transition is discussed.
\end{abstract}


\maketitle
\section{Introduction} 
Measurement in quantum many-body systems leads to nontrivial dynamics and produces exotic phases. 
Measurement of entangled resource states induces measurement-based quantum computation \cite{Raussendorf2001,Briegel2009} and produces long-range entanglement states or topological order \cite{Tantivasadakarn2022}. 
With a suitable choice of measurement types, measurement-only circuits generate unconventional phases of matter such as symmetry-protected topological (SPT) phases \cite{Lavasani2021,Klocke2022,KI2023}, topological orders \cite{Lavasani2021_2,Negari2023}, and non-trivial thermal and critical phases \cite{Ippoliti2021,Sriram2023,KOI2023,Lavasani2023,Zhu2023}. 
As a further example, hybrid random-unitary circuits exhibit measurement-induced entanglement phase transitions \cite{Li2018,Skinner2019,Li2019,Vasseur2019,Chan2019,Szyniszewski2019,Choi2020,Bao2020,Gullans2020,Jian2020,Zabalo2020,Sang2021,Sang2021_v2,Nahum2021,Sharma2022,Fisher2022_rev,Block2022,Liu,Richter2023,Sierant2023,Kumer2023}. 

\begin{figure}[b]
\begin{center} 
\includegraphics[width=7cm]{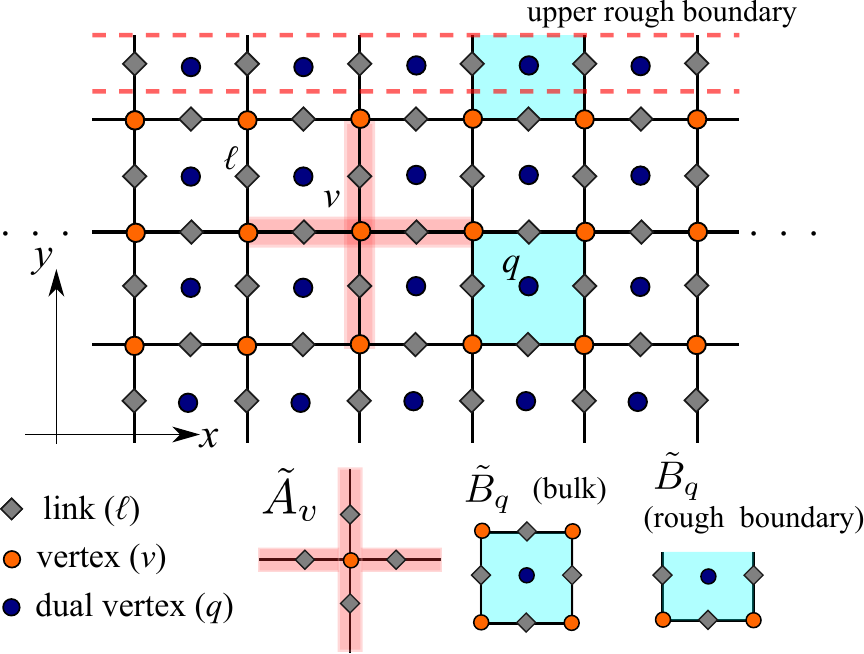}  
\end{center} 
\caption{Schematic figure of cylinder system with rough boundaries. This figure displays only the upper rough boundary. 
The star operator (the red shaded object) and plaquette operator (right-blue shaded object) are considered.  
The red dashed regime is an upper rough boundary regime.}
\label{Fig1}
\end{figure}

Recently, toric code \cite{Kitaev2003} was experimentally realized \cite{Google_quantum_AI}. 
Target geometry of the system is torus, providing two-qubit quantum memory \cite{Kitaev2003,Dennis2001,Wang2003}. 
Recent studies, however, showed that the interplay of an open geometry, such as the cylinder with rough boundary (Fig.~\ref{Fig1}), and measurement can give rich physical phenomena \cite{Verresen2022,Wildeboer2022,Negari2023}.
This observation was obtained from the viewpoint of lattice gauge theory \cite{Kogut1979}. 
In quantum information aspect, edge mode on the boundary works as a qubit \cite{Verresen2022,Wildeboer2022,KI2023_2}, which is called subsystem code~\cite{Poulin2005,Bacon2006}.
From the gauge theory side, it was recently suggested that edge mode creates some kind of long-range order (LRO) and it distinguishes the Higgs and confinement phases, although it was believed for a long time that the two phases are adiabatically connected without any transitions \cite{Fradkin1979}. 
This LRO indicates that the Higgs phase is a SPT phase protected by a higher-form symmetry \cite{Verresen2022}, i.e., the bulk SPT order induces  the boundary LRO. 
This developments attract lots of attention these days. 

The above discussion was given for Hamiltonian systems. 
In this work, we shall study a closely-related system with the cylinder geometry by using a measurement-only scheme.
In particular, we investigate the emergence of edge modes and transition from the deconfined (toric code) to Higgs phases as the measurement corresponding to `Higgs coupling' is getting frequent. 
Our protocol is as follows:
We prepare a toric code state as the initial state, and observe what happens on performing local projective measurement \textit{on the bulk}. 
The reason for performing measurement only on the bulk is that the interplay of a non-local gauge-invariant operator (NLGIO) symmetry and Higgs condensation in the bulk induces LRO with spontaneous-symmetry breaking (SSB) on boundary, a signature of the SPT.
In fact, by numerical methods, we verify the emergence of the boundary LRO as a clear measurement-induced phase transition on the boundary. 
Furthermore, we find some noticeable features of the transition, and discuss that it is evidence of 
the interplay of the NLGIO and Higgs condensation.

The rest of this paper is organized as follows. 
In Sec.~II, we introduce measurement protocol on toric code state. 
In Sec.~III, Hamiltonian systems corresponding to our measurement protocol are explained. 
In particular, we discuss the connection of a lattice-gauge-Higgs model and toric code model. 
In Sec.~IV, we search a boundary phase transition by employing efficient numerical methods
in the stabilizer formalism and discuss physical properties and indications of the phase transition rather in detail.
Then, we study the bulk properties of the measured system in Sec. V. 
In particular, we investigate the phase transition in the bulk and discuss its relationship to the boundary transition.
Section VI is devoted for conclusion.

\section{Measured toric code system}
We consider a code system composed of $L_x \times L_y$ plaquettes ($q$-lattice) and $L_x \times (L_y-1)$ vortices ($v$-lattice).
Physical qubits reside on links of the $v$-lattice, and cylinder geometry and rough boundaries in the $y$ direction are employed. 
The schematic image of the system is shown in Fig.~\ref{Fig1}.
In our protocol, we consider a stabilizer state specified by a set of stabilizer generators \cite{Nielsen_Chuang} focusing on pure states, 
$S_{\rm int}=\{\tilde{A}_v | v\in \mbox{all} \ v \} + \{ \tilde{B}_q|q\in 
\mbox{all plaquette but} \ q_0 \} + \{S^u_x \}$,
where $q_0$ is one of single $q$-lattice sites to avoid the redundancy to make all elements of the stabilizer generator linearly independent.
In the above, $\tilde{A}_v$ and $\tilde{B}_q$ are the star and plaquette operators of the toric code, defined by $\tilde{A}_v=\prod_{\ell_{v} \in v} \sigma^x_{\ell_{v}}$ and $\tilde{B}_q=\prod_{\ell_{q} \in q} \sigma^z_{\ell_{q}}$, 
where $\ell_{v} \in v$ stands for links emanating from vertex $v$, and $\ell_{q} \in q$ for links composing plaquette $q$ and 
$\displaystyle{S^{u}_x= \prod_{\ell\in {\rm URB}}\sigma^x_{\ell}}$, (URB stands for the upper rough boundary \cite{LRB}), which is a non-local stabilizer related to the topological-symmetry generators \cite{Wildeboer2022}. 
Physically, $\displaystyle{S^{u}_x}$ is nothing but the spin flip operator on the upper rough boundary. 
This stabilizer state corresponds to one of the toric code ground states, the Hamiltonian of which is shown shortly.

We use this stabilizer state as an initial state and perform local $\sigma^z$-projective measurement for each qubit with probability $p$ \textit{except for the dangling links on the upper and lower rough boundaries}.
We call measurement pattern sample, and we average physical quantities over samples.
Practically, under the process of projective measurement in the stabilizer formalism \cite{Gottesman1997,Nielsen_Chuang}, $\sigma^z$-projective measurement at link $\ell$ removes one of the star operators $\tilde{A}_v$ residing on the boundary vertices of $\ell$ ($v_1$ and $v_2$), and then $\sigma^z_\ell$ becomes a stabilizer as well as the product of the star operators $(\tilde{A}_{v_1}\tilde{A}_{v_2})$. 
That is, the initial toric code state tends to lose $\tilde{A}_v$'s depending on the probability $p$, and the number of $\sigma^z$-stabilizer generator increases instead. It should be also remarked that $(\tilde{A}_{v_1}\tilde{A}_{v_2})$ loses $\sigma^x_\ell$ on the $\ell$ meeting $v_1$ and $v_2$, and has six $\sigma^x$'s.
This plays an important role for discussion on the bulk transition.

Numerically, the measurement process is performed by the efficient numerical simulation by the stabilizer formalism \cite{Gottesman1997,Aaronson2004}. 
Throughout this study, we ignore the sign and imaginary factors of stabilizers, which give no effects on observables that we focus on.
Note that the order of the local measurement is interchangeable since all measurement operators $\{\sigma^{z}_{\ell}\}$ are commutative with each other. 

\section{View point from Hamiltonian formalism} 
Before going to the numerical analysis, we show that our measurement protocol 
is to be compared with the toric code in magnetic fields on cylinder, Hamiltonian of which is given by 
\begin{eqnarray}
H_{\rm TC}&=&\sum_{v}h_x\tilde{A}_v +\sum_{q}h_z\tilde{B}_q+\sum_{\ell \notin \mbox{rough}}J_z\sigma^z_{\ell}
+\sum_{\ell}J_x\sigma^x_{\ell}.\nonumber\\
\label{HTC}
\end{eqnarray}
In this section, we briefly review the lattice gauge-Higgs model and its physical properties, recently investigated from novel point of view \cite{Verresen2022}, and also introduce  NLGIO symmetries, 
which play an important role for the interpretation of our numerical results in the following sections.
We also explain that the toric code model [Eq. (\ref{HTC})]  is a gauge fixing version of the lattice gauge-Higgs model.

\subsection{Lattice-gauge-Higgs model}
We shall explain the relation of the toric code and gauge-Higgs model in two spatial dimensions, especially under cylinder geometry. 
Accessible reviews of the lattice gauge theory (LGT) are Refs.~\cite{Kogut1979,IchinoseMatsui2014}. 

\begin{figure}[b]
\begin{center} 
\includegraphics[width=7.5cm]{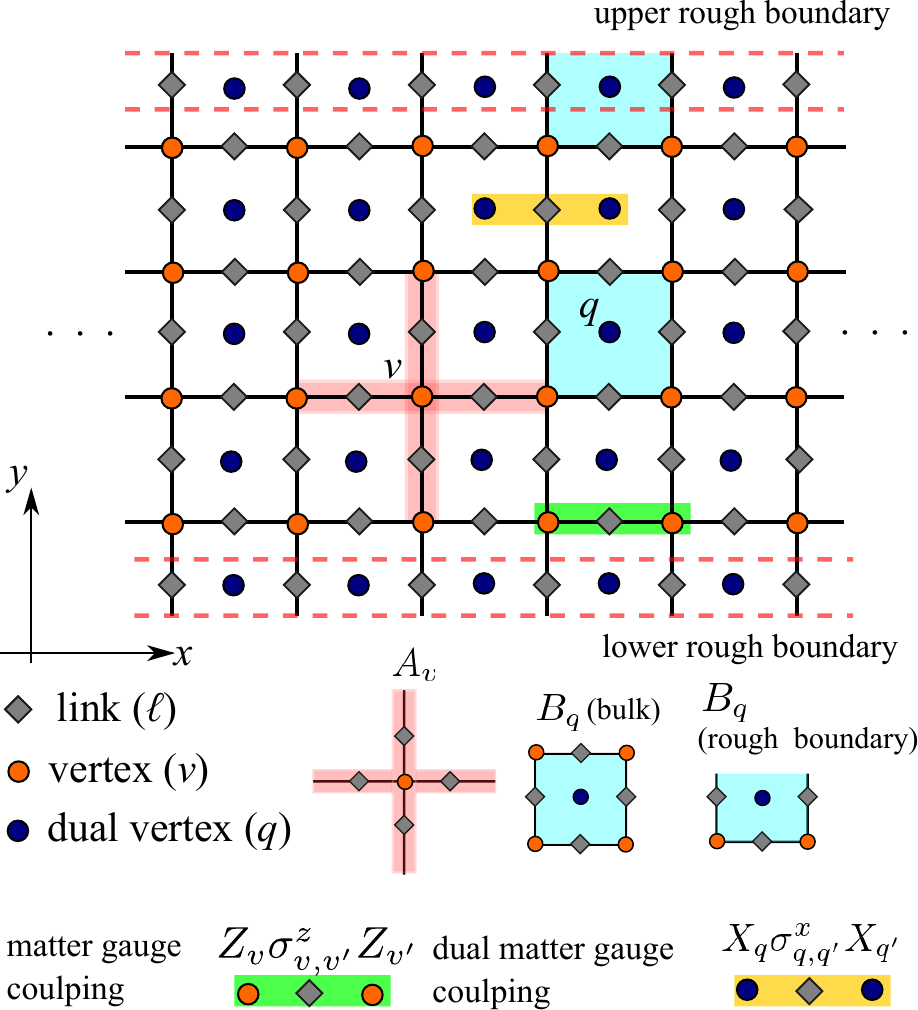}  
\end{center} 
\caption{Schematic figure of an extended lattice gauge-Higgs model with rough boundaries. The lattice in the $x$-direction is periodic and in $y$-direction is open. The red dashed lines indicate the regime upper and lower boundaries}
\label{FigA1}
\end{figure}

The ancestor model of $H_{\rm TC}$ is a (2+1)-D $Z_2$ extended gauge-Higgs model, Hamiltonian of which is given by
\begin{eqnarray}
H_{\rm GHM}&=&\sum_{v}h_x X_v 
+\sum_{(v,v')}J_zZ_{v}\sigma^z_{v,v'}Z_{v'}\nonumber\\
&+&\sum_{q}h_zZ_q+\sum_{(q,q')}J_xX_{q}\sigma^x_{q,q'}X_{q'},
\label{HGHM}
\end{eqnarray}
where $X$ and $Z$ are Pauli operators defined on $q$- and $v$-lattice sites. 
The first and second terms are a $Z_2$ matter chemical potential on vertices and its gauge coupling term, the third and fourth terms are $Z_2$ matter chemical potential on plaquette sites and its gauge coupling term. 
The schematic image of the system and each terms are shown in Fig.~\ref{FigA1}.

Here, the two gauge-invariant conditions, the Gauss laws, for the physical subspace, are given by
\begin{eqnarray}
A_v|\psi\rangle=|\psi\rangle,\:\:
B_q|\psi\rangle=|\psi\rangle.
\label{const}
\end{eqnarray}
where
\begin{eqnarray}
&&A_v=X_v\prod_{\ell_{v} \in v} \sigma^x_{\ell_{v}}\equiv X_v\tilde{A}_v, \nonumber  \\
&&B_q=Z_q\prod_{\ell_{q} \in q} \sigma^z_{\ell_{q}}\equiv Z_q\tilde{B}_q,
\label{GvBp}
\end{eqnarray}
and $\ell_{v} \in v$ stands for links emanating from vertex (site) $v$, and 
$\ell_{q} \in q$ for links composing plaquette (box) $q$.
The $Z_2$-electric matter is defined on each vertex $v$, $(X_v,Z_v)$, and its magnetic dual,  $(X_q,Z_q)$, on each dual vertex $q$ 
(i.e., plaquette of the original lattice). 
On the other hand, the $Z_2$ gauge field is defined on links and denoted by $(\sigma^x_\ell, \sigma^z_\ell)$, 
$\sigma^z_{v,v'}$ denote a gauge variable on link connecting  neighboring vertices $v$ and $v'$, 
and $\sigma^x_{q,q'}$ denote a gauge variable on link connecting neighboring dual vertices $q$ and $q'$. 

There are two differences between the system given by $H_{\rm GHM}$ [Eq.~(\ref{HGHM})] and the ordinary 
$Z_2$ gauge-Higgs LGT \cite{Fradkin1979}: 
\begin{enumerate}
\item[(I)]
Dual matter field couples with the gauge field and the coupling term, $X_{q}\sigma^x_{q,q'}X_{q'}$ 
is added to the Hamiltonian besides the ordinary $Z_2$ electric matter-gauge coupling.
This $Z_2$ degrees of freedom residing on each plaquette $q$, $(X_p,Z_p)$, corresponds to `particle' carrying 
magnetic flux (magnetic charge), and its hopping induces fluctuation of the gauge field and, therefore, confinement of electric charges. This degree of freedom is often called \textit{m-anyon}, whereas the other one \textit{e-anyon}.
\item[(II)]
The model has an an additional local gauge symmetry by the presence of the magnetic-charge degrees of freedom, 
\begin{equation}
X_q \rightarrow X_q V_q, \; \; 
\sigma^x_{q,q'} \rightarrow  V_q\sigma^x_{q,q'}V_{q'},  \;\; V_q, V_{q'} \in Z_2,
\label{second}
\end{equation}
and we impose additional Gauss-law constraint on the physical state, Eqs.~(\ref{const}) and (\ref{GvBp}).
One may wonder that the system (\ref{HGHM}) reduces to the ordinary gauge-Higgs model by 
`integrating out' the magnetic charge degrees of freedom via, e.g., employing unitary gauge of the second local gauge symmetry in Eq.~(\ref{second}).
In fact as we show shortly, disentangling of electric and magnetic particles generates the ordinary 
Hamiltonian of the gauge-Higgs model in unitary gauge.
\end{enumerate}

Here, we comment that on introducing rough and smooth boundaries for the above Hamiltonian, the model describes subsystem quantum code producing fault-tolerant qubit as proposed in Ref.~\cite{Wildeboer2022}, although the explicit form of $H_{\rm GHM}$ in Eq.~(\ref{HGHM}) was not shown there. 
There, each term of the Hamiltonian (\ref{HGHM}) is categorized as ``gauge'' operator in the subsystem-code literature \cite{Poulin2005}. 

\subsection{Symmetry and relation to Toric code}
The model $H_{\rm GHM}$ with the above open boundary conditions is known to have two important global topological symmetries \cite{Wildeboer2022}, generators of which are given by,
\begin{eqnarray}
P&=&\prod_{v}X_v,\:\:
S_Z=\prod_{q}Z_q,
\label{symmetry1}
\end{eqnarray}
$P$ is the parity of the total electric charge, corresponding to the global spin flip on each vertex,
and similarly, $S_Z$ is the parity of the total magnetic flux per plaquette.

Furthermore as shown in Ref.~\cite{Verresen2022}, there are two symmetries, besides the global symmetries 
in Eq. (\ref{symmetry1}),
\begin{eqnarray}
W_{\gamma}&=&\prod_{\ell \in \gamma}\sigma^z_{\ell},\:\:
H_{\gamma}=\prod_{\ell \in \gamma}\sigma^x_{\ell}.
\label{symmetry2}
\end{eqnarray}
$W_{\gamma}$ and $H_{\gamma}$ are called one-form symmetries, which have been extensively studied recently \cite{Gaiotto2015, McGreevy2022}.
Although the $J_x$ and $J_z$ terms in the Hamiltonian $H_{\rm GHM}$ [Eq.~(\ref{HGHM})] explicitly break 
the one-form symmetries, it was shown that the higher-form symmetry is generally robust and 
give non-trivial effect on dynamics of the system \cite{Verresen2022}. 
In the $(2+1)$-D system, $H_\gamma$ can be regarded as 't Hooft loop (string) dual to Wilson loop (string) $W_\gamma$.
In particular, the path $\gamma$ in the one-form symmetry $W_{\gamma}$ ($H_{\gamma}$) is arbitrary.
It is easily verified that by using the Gauss laws in Eq.~(\ref{GvBp}), 
$P$ is expressed as a 't Hooft `loop' residing on the upper and lower rough boundaries. 
From the view points of these symmetries, the Higgs phase was discussed in \cite{Verresen2022}, showing that the Higgs phase is SPT phase from the perspective of these symmetries.
In this study, however, we focus on related but slightly different symmetry operators LGIO's as introduced in Sec.III D.
\begin{figure}[t]
\begin{center} 
\includegraphics[width=7.5cm]{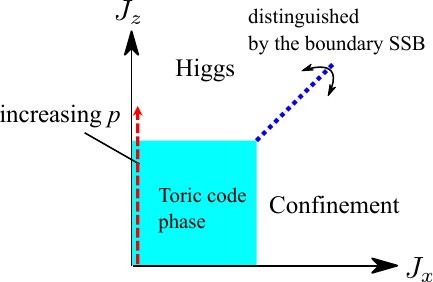}  
\end{center} 
\caption{Schematic image of phase diagram of the Hamiltonian $H_{\rm TC}$ on the cylinder geometry. We assume $h_x$ and $h_z$ are some constant. The dashed arrow represents our target measurement protocol related to the parameter sweep in the Hamiltonian formalism.}
\label{FigA1-2}
\end{figure}

Now we show that the model $H_{\rm GHM}$ is related to a generalized toric code $H_{\rm TC}$. 
We consider the following unitary transformations \cite{Wildeboer2022,Verresen2022}:
\begin{eqnarray}
U_v&=&H\biggl(\prod_{v}\prod_{\ell\in v} (CZ)_{v,\ell}\biggr)H,\\
U_q&=&H\biggl(\prod_{q}\prod_{\ell\in q} (CZ)_{q,\ell}\biggr)H,
\end{eqnarray}
where $H$ is the Hadamard transformation on each link and 
$(CZ)_{i,j}$ is a controlled $Z$-gate for the site $i$ and link $j$.
Applying the above transformation to $H_{\rm GHM}$, we obtain the following effective disentangled model
\begin{eqnarray}
&&U_vU_qH_{\rm GHM}(U_vU_q)^\dagger\equiv H_{\rm TC},\nonumber
\end{eqnarray}
The above unitary disentanglement corresponds to gauge fixing with unitary gauge. 
The Hamiltonian $H_{\rm GHM}$ is related to the Hamiltonian $H_{\rm TC}$.

\subsection{Ground state properties}
The gauge-Higgs model on infinite system was analyzed in Fradkin and Shenker \cite{Fradkin1979}, 
and its phase diagram has been investigated in detail \cite{Trebst2007,Vidal2009,Tupitsyn2010,Dusuel2011,Wu2012,Xu2023}.
There are three phases, deconfined (toric code), Higgs, and confined phases, but with periodic boundary conditions, the Higgs and confinement phases are connected without  any thermodynamic singularities \cite{Fradkin1979}. 
The schematic image is shown in Fig.~\ref{FigA1-2}. 
In particular, for the spatial dimension $D\geq 2$, the Higgs phase exists for $|J_z|> |J_x|,|h_x|,|h_z|$, as the $J_z$-term is nothing but the hopping of the charged particle. 

When the cylinder geometry is employed, the Higgs and confinement phases can be distinguished. 
As suggested in Ref.~\cite{Verresen2022}, in the deep Higgs phase ($J_z\to \infty$), $\langle \sigma^z_{\ell}\rangle=-1$ in the bulk, then the Hamiltonian $H_{\rm TC}$ can be reduced to a $1D$ transverse field Ising model on the rough boundary,
\begin{eqnarray}
H_{rb}=\sum_{\langle\ell_x,\ell'_{x}\rangle}h_z\sigma^z_{\ell_{x}}\sigma^z_{\ell'_{x}}+\sum_{\ell_x}J_x\sigma^x_{\ell_x},
\label{EH1}
\end{eqnarray}
where $\ell_x$ is a dangling link on the (upper or lower) rough boundary, $\langle\ell_x,\ell'_{x}\rangle$ represents nearest-neighbor pairs of the dangling links on the rough boundary.
This Hamiltonian $H_{rb}$ exhibits the SSB for $|h_z|>|J_x|$. 
Thus, the Higgs phase can be distinguished from the other two phases by observing the boundary SSB.

Here, we further comment on the ground-state property of the model $H_{\rm TC}$ on the cylinder.
For $J_x=J_z=0$ (toric code phase), the model $H_{\rm TC}$ is exactly solvable but its state degeneracy is different from the system with the conventional torus geometry. 
From the stabilizer viewpoint \cite{Nielsen_Chuang}, we count the number of independent stabilizers.
Here, as to the plaquette operator $\tilde{B}_q$, one constraint exists $\prod_{q\in all}B_q=1$, thus there are $L_xL_y-1$ independent stabilizers of $\tilde{B}_{q}$, denoted by a set of generator of stabilizer, 
$\{ \tilde{B}_q|q\in 
\mbox{all plaquette but} \ q_0 \}$, where $q_0$ is one of single $q$-lattice sites to avoid the redundancy. 
On the other hand, as to the star operator $\tilde{A}_v$, there is 
$L_x(L_y-1)$ independent stabilizer generators since $\prod_{v\in all}\tilde{A}_v\neq 1$. 
From the above counts,  totally $2L_xL_y-L_x-1$ linear independent stabilizers exist and from the fact that the total number of link qubits is $L_x(2L_y-1)$, the degeneracy is $2^{(L_x(2L_y-1))-(2L_xL_y-L_x-1)}=2$. 
That is, two-fold degeneracy appears for the entire energy levels. 

In the exact toric code phase for  $J_x=J_z=0$, two operators $\prod_{\rm periodic}\sigma^x_\ell$ and $\prod_{\rm cross}\sigma^z_\ell$ commute with $H_{\rm TC}$ and anti-commute with each other, where `periodic' denotes a closed loop on the dual lattice in the periodic direction ($x$-direction) of the cylinder and `cross' a string crossing the cylinder from the upper to lower rough boundaries.
The two-fold degeneracy can be understand by the existence of them, and therefore it is expected that
this degeneracy is robust against local perturbations with finite $J_x$ and $J_z$ as long as the system is in the toric-code phase.

Furthermore, we focus on the two-fold degenerate ground states of $H_{\rm TC}$ with $J_x=J_z=0$ in the toric code phase denoted by $|\psi_{G1}\rangle$ and $|\psi_{G2}\rangle$. 
These states can be distinguished by a string operator $S^u_x$  introduced in Sec. II as $S^u_x|\psi_{G1(2)}\rangle =+1(-1)|\psi_{G1(2)}\rangle$.
The operator $S^u_x$ is linearly independent of all $\tilde{A}_v$'s and $\tilde{B}_q$'s and also commute with them, thus it can be a linearly independent additional stabilizer. 
In our protocol with projective measurement, we choose the state $|\psi_{G1}\rangle$ as the initial state.

\subsection{NLGIO symmetry in Toric code}
As important symmetries, we consider NLGIO for $H_{\rm TC}$ with $J_x=0$. 
Two types of them are:
(I) $G_{lo,1}\equiv \prod_{\ell\in\Gamma_0} \sigma^z_{\ell}$
with an arbitrary close loop $\Gamma_0$, 
(II) $G_{lo,2}=\sigma^z_{\ell_{r1}}\biggl[\prod_{\ell\in\Gamma_b}\sigma^z_{\ell}\biggr]\sigma^z_{\ell_{r2}}$ where $\ell_{r1}$ and $\ell_{r2}$ are two arbitrary dangling links on the upper rough boundary and the string in the bulk $\Gamma_b$ connecting the $\ell_{r1}$ and $\ell_{r2}$ links. 
Both these operators satisfy $[H_{\rm TC},G_{lo,1(2)}]=0$ for $J_x=0$. 
In particular, the second-type NLGIO $G_{lo,2}$ plays an important role in the subsequent study.
Another conserved charge is $\prod_{\ell \in {\rm rough}}\sigma^x_\ell$, which is nothing but the global charge parity of the gauge-Higgs model, and is an element of stabilizers of the initial state. 

We note that the NLGIO symmetries are different from the magnetic-one-form symmetry considered in \cite{Verresen2022}, etc.
They are exact symmetries even in the presence of the $A_v$-terms in $H_{\rm TC}$.
From this fact, both the NLGIO's are elements of the stabilizer group and can provide a useful picture for understanding the relationship between the bulk Higgs phase and boundary order, as we explain later on.

\begin{figure}[t]
\begin{center} 
\includegraphics[width=8cm]{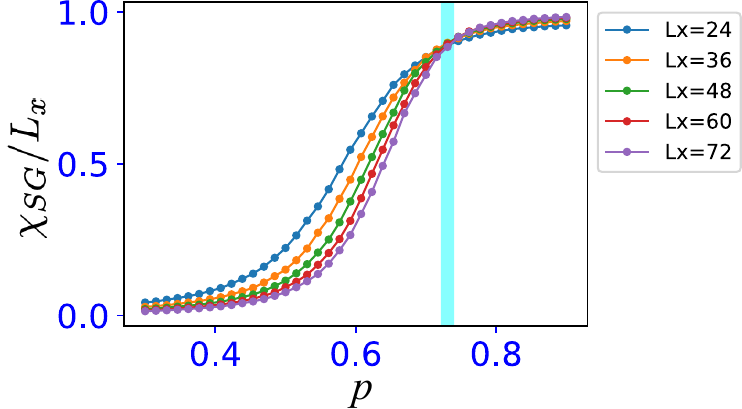}  
\end{center} 
\caption{Behavior of the average value of $\chi_{SG}/L_x$.
The right-blue shaded line represents the crossing point of the different system size data.
We used $3\times 10^3$ samples. We fix $L_y=6$.}
\label{FigA2}
\end{figure}
\section{search for a phase transition on the boundary}

In this section, we study `phase diagram' of  the measured toric code. 
The initial state employed in the present numerical study is one of the exact ground states stabilized by $S_{\rm int}$.
Key point of this study is to observe edge physics on the rough boundaries. 
As mentioned in the previous section, the model $H_{\rm TC}$ reduces to an effective rough boundary Hamiltonian similar to the transverse field Ising model. 
Thus, it is expected that SSB of $Z_2$-full spin-flip symmetry ($S^{u}_x$) with a magnetic LRO emerges
as the strength of the $\sigma^z$-term increases in $H_{\rm TC}$ (see Fig.~\ref{FigA1-2}).
However, we note that this expectation is \textit {not} obvious as spins on the boundaries keep intact on measurement.

We shall investigate the above qualitative expectation in our measurement protocol by the efficient stabilizer numerical methods. Performing local $\sigma^z$-projective measurement for each qubit with probability $p$ can correspond to increasing $J_z$ with $J_x=0$ in the Hamiltonian $H_{\rm TC}$ as shown in Fig.~\ref{FigA1-2}.

We study numerically the edge states on the rough boundaries emerging with LRO. 
We are interested in quasi-one-dimensional subsystem of upper or lower boundary.
Thus, our numerics prioritize to vary $L_x$ with $L_y$ fixed. 
Increasing both $L_x$ and $L_y$ (especially $L_y>4$) requires large computational cost being inaccessible. 
We mostly set $L_y=6$ with care since for too small $L_y$, transition-like behavior on the boundary disappears.

We introduce the Edward-Anderson spin-glass (SG) order parameter \cite{SG_ex} defined by
$$
\chi_{SG}=\frac{1}{L_x}\sum_{\ell_x,\ell'_x}C_{SG}(\ell_x,\ell'_x)$$ with 
$$
C_{SG}(\ell_x,\ell'_x)=\langle \psi|\sigma^z_{\ell_x}\sigma^z_{\ell'_x}|\psi\rangle^2-\langle \psi|\sigma^z_{\ell_x}|\psi\rangle^2 \langle \psi|\sigma^z_{\ell'_x}|\psi\rangle^2,
$$
where $|\psi\rangle$ is the stabilizer state and $\ell_x$'s are links on the boundary. 
As a target observable, we observe the variance of $\chi_{SG}$ divided by $L_x$, 
$F_v(p,L_x)\equiv \mbox{var}(\chi_{SG})/L_x$. 
This quantity measures sample-to-sample fluctuations and is useful to detect a phase transition on the boundary. 
$F_v(p,L_x)$ shows a clear peak at a transition point \cite{Kjall2014}.

\begin{figure}[t]
\begin{center} 
\includegraphics[width=7.5cm]{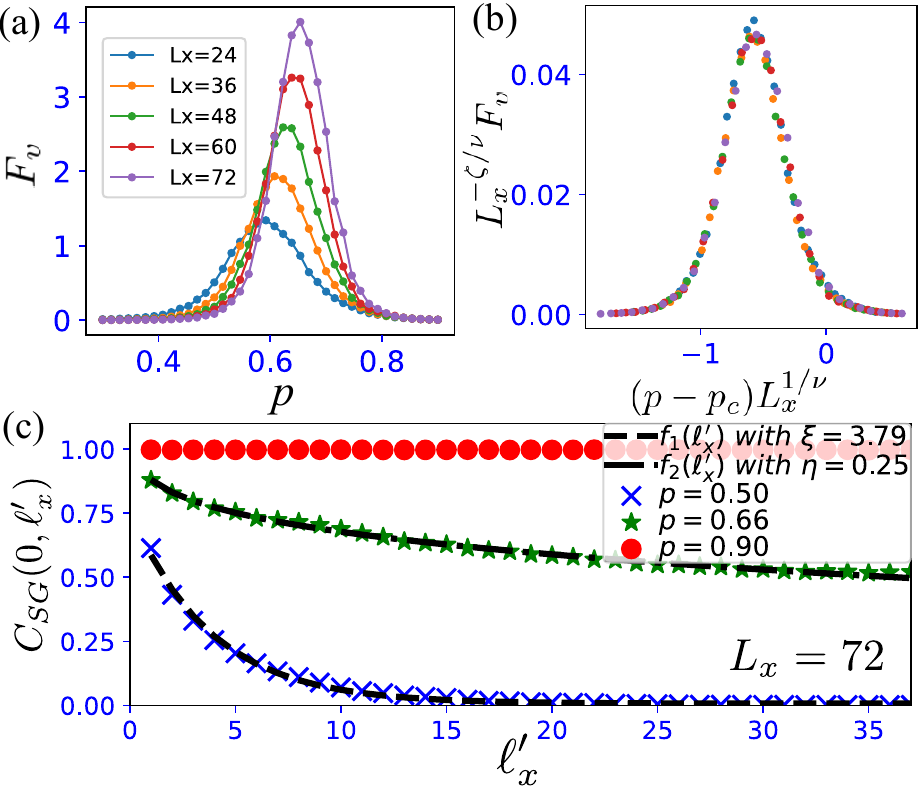}  
\end{center} 
\caption{(a) System-size dependence of $F_v$. 
(b) Scaling function. 
(c) Correlation for $C_{SG}(\ell_x,\ell'_x)$.
Here, we chose $\ell_x=0$, $L_x=72$. We used the following fitting functions:
$f_1(\ell_x')=a\exp(-\ell_x'/\xi)+b$ and $f_2(\ell_x')=a\ell_x'^\eta+b$, where $a$, $b$, $\xi$, and $\eta$ are fitting parameters.
Scatter plots represent numerical results specified by probability $p$. 
Note that peak of $F_v$ in $L_x=72$ system is located at $p=0.66$. 
Thus, physical quantities are expected to show critical behavior at $p=0.66$ in $L_x=72$ system. 
We used $(3-4)\times 10^3$ samples for (a)-(c).
}
\label{Fig2}
\end{figure}
In Fig.~\ref{FigA2}, we first display the calculations of $\chi_{SG}/L_x$ as a function of $p$ for various $L_x$, which exhibit smooth curves crossing with each other at $p\sim 0.75$. 
This result implies the existence of a phase transition there. 

Next, figure \ref{Fig2}(a) shows calculations of $F_v(p,L_x)$.
We find that the peak grows with increasing $L_x$ and the peak shifts toward larger $p$ with increasing $L_x$. 
This is a typical continuous phase transition behavior.
This result indicates the existence of the boundary phase transition. 
As $F_v(p,L_x)$ shows, the fluctuations of the SG order gets simultaneously zero for all system sizes
 at $p\sim 0.75$, and solid SG order appears there in the thermodynamic limit.

The above phenomenon can be understood by the NLGIO symmetry $G_{lo,2}$ of $H_{\rm TC}$ and condensation of Higgs. 
As we show later on, the SSB of the NLGIO symmetry $G_{lo,2}$ is \textit{restored} in this regime.
The NLGIO symmetry $G_{lo,2}$ dictates for any states under consideration $G_{lo,2}|\psi\rangle=\sigma^z_{\ell_{r1}}\biggl[\prod_{\ell\in\Gamma_b}\sigma^z_{\ell}\biggr]\sigma^z_{\ell_{r2}}|\psi\rangle \propto |\psi\rangle$.
If the condensation of Higgs boson takes place for length scale $|\Gamma_b|$, $\langle \prod_{\Gamma_b} \sigma^z_\ell\rangle\neq 0$, we have $\langle \sigma^z_{\ell_{r1}}\sigma^z_{\ell_{r2}}\rangle\neq 0$.
Therefore, the above numerical result indicates that perfect Higgs condensation takes place at $p\sim 0.75$. 
Here, the physical picture is that the proliferation of the bulk Higgs condensate induces condensation of NLGIO with various shapes connecting two different links on the boundary. 
This condensation of  string operators (corresponding to Wilson string with both of the edges attached to the boundary) induces long-range order $\langle \sigma^z_{\ell_{r1}}\sigma^z_{\ell_{r2}}\rangle\neq 0$. 

In an intuitive picture, as $G_{lo,2}$ is an element of the stabilizer group, the finite Higgs condensation $\langle \prod_{\Gamma_b} \sigma^z_\ell\rangle\neq 0$ generates an effective stabilizer such as $\sigma^z_{\ell_{r1}}\sigma^z_{\ell_{r2}}$.
In this sense, effective interactions on the boundary are induced by the bulk Higgs condensation.
We expect that as the above string $\Gamma_b$ in the bulk is arbitrary, the critical value $p$ can decrease for larger $L_y$ as the number of strings $\Gamma_b$'s participating the effective boundary interactions increases.
Similar mechanism works for enhancing the Higgs condensation for larger $L_y$.

We identify the transition point and its criticality by applying the scaling analysis to $F_v(p,L_x)$ by means of  pyfssa numerical package~\cite{Pyfssa1,Pyfssa2}. 
Here, the scaling ansatz is set as $F_v(p, L_x)=L_x^{\frac{\zeta}{\nu}}\Psi((p-p_c)L_x^{\frac{1}{\nu}})$, where $\Psi$ is a scaling function, $\zeta$ and $\nu$ are critical exponents and $p_c$ is a critical transition probability in the thermodynamics limit. 
We observe the clear data collapse as shown in Fig.~\ref{Fig2}(b). 
Here, the transition probability is estimated by $p_c=0.757 \pm 0.035$ and the exponents, $\nu = 2.54\pm 0.52$ and $\zeta=2.64\pm 0.34$. 
The above scaling analysis indicates almost perfect SG order on the rough boundary for $p\geq 0.757$ for all system sizes, including infinite $L_x$. 
We note that $p_c$ corresponds to the scaling function $x=0$ in Fig.~\ref{Fig2} (b). The point $x=0$ in the scaling function is off-peak. 
This implies that the ordered phase (SSB phase) emerges after strict suppression of the fluctuations of the spins on the boundary in the thermodynamic limit.
This is some peculiar feature of the present system, which comes from the fact that effective couplings between boundary spins are induced by the bulk measurement and they do not exist at the onset. 

\begin{figure}[t]
\begin{center} 
\includegraphics[width=8cm]{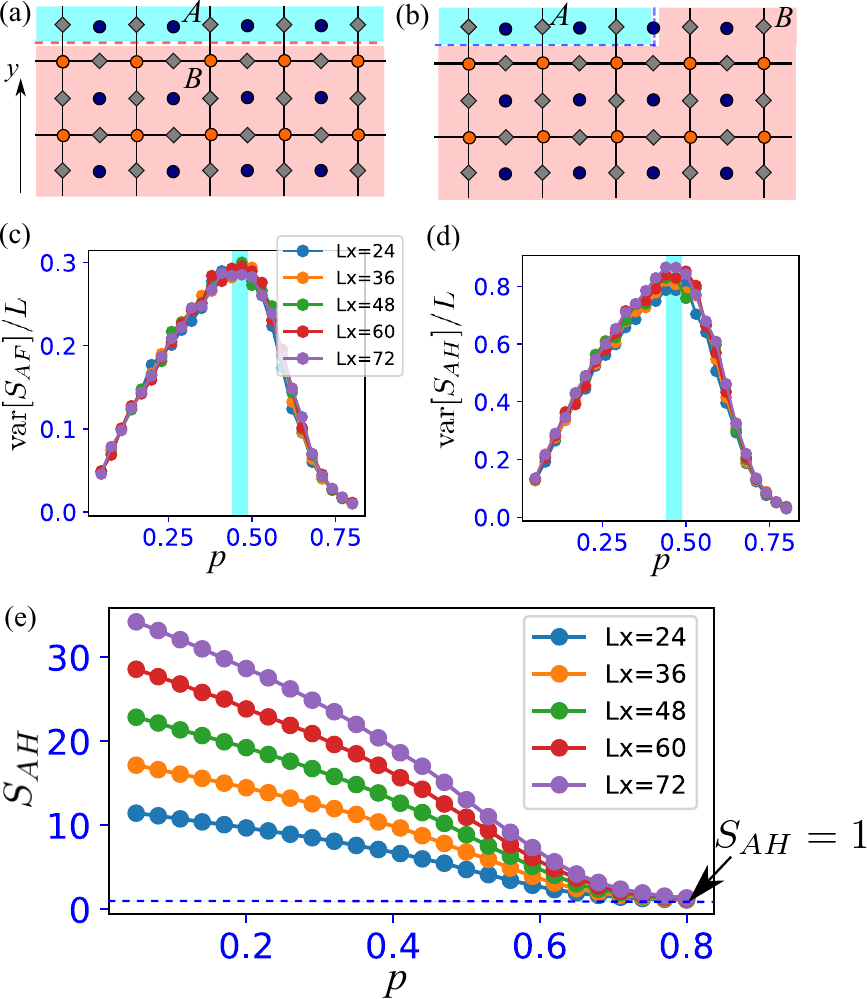}  
\end{center} 
\caption{Partition for $S_{AF}$ [(a)] and $S_{AH}$ [(b)]. 
Sample-to-sample variance for $S_{AF}$[(c)] and $S_{AH}$ [(d)]. (e) System-size dependence of $S_{AH}$, where we averaged over all measurement samples. We used $3\times 10^3$ samples for (c), (d) and (e).}
\label{Fig3}
\end{figure}
For further study, we observe the correlation function $C_{SG}(\ell_x,\ell'_x)$ for the $L_x=72$ and $L_y=6$ system in the vicinity of  the peak of $F_v(p,L_x)$, which we denote  $p=p'_c$.
The results are displayed in Fig.~\ref{Fig2} (c). 
For $p>p'_c$, a clear long-range SG order exists, as expected from the above observation. 
On the critical point $p=p'_c$, the correlator is well-fitted by the power-law decay, estimated as $\propto |\ell_x-\ell_x'|^{\eta}$ with $\eta\sim 0.25$.
Finally for $p < p'_c$, the result is well-fitted by the exponential decay, estimated as $\propto \exp[-|\ell_x-\ell_x'|/\xi]$ with $\xi\sim3.79$, indicating a disordered phase.
The analysis verifies the SG boundary phase that is induced by the bulk measurement enhancing Higgs condensation.
In passing, we examine another fitting for $C_{SG}(\ell_x,\ell'_x)$ at the criticality such as 
$e^{-\alpha|\ell_x-\ell_{x'}|^\beta}$~\cite{Fisher1995,Fisher1999}.
However, we find that the power-law fitting is better than that. 
We also observed effects of the system size in the $y$-direction, $L_y$, to the rough-boundary phase transition. 
The additional data is shown in Appendix, the result of which is consistent with the consideration by the Hamiltonian formalism.

We further investigate entanglement properties of our system. 
We focus on entanglement entropy (EE) for a subsystem $A$ (its complement is subsystem $B$).
(We use logarithm with base $2$ for EE.) 
Here, we consider two different partitions shown in Fig.~\ref{Fig3}(a) and ~\ref{Fig3}(b). 
First partition is that the subsystem $A$ is the upper rough boundary and its EE is denoted by $S_{AF}$.
In the second one, the subsystem $A$ is the half of the upper rough boundary and its EE is denoted by $S_{AH}$.   
The EE can be calculated from the number of linearly-independent stabilizers within a target subsystem $A$ and the number of qubit of the subsystem $A$ \cite{Fattal2004,Nahum2017}. 

The sample-to-sample variance of $S_{AF}$ divided by $L_x$, $\mbox{var}[S_{AF}]/L_x$, is plotted in Fig.~\ref{Fig3}(c) to shed light on a bulk transition behavior motivated by Ref.~\cite{Kjall2014}. We find the peak and no system-size dependence indicating a crossover. 
The EE $S_{AF}$ evaluates the degree of separation between the bulk and the upper rough boundary. 
From the data, the rough boundary is separated from the bulk for $p\gtrsim 0.45$, smaller than the rough boundary transition point $p_c$. 
This behavior gives an intuitive scenario that after the entanglement separation between the rough boundary and bulk, the SG LRO on the rough boundary starts to develop. This result is consistent with the fact that the proliferation of Higgs condensation in the bulk 
enhances short-range entanglement and hinders long-range one (i.e., the topological order).

We next observe the sample-to-sample variance of $S_{AH}$ divided by $L_x$, $\mbox{var}[S_{AH}]/L_x$ plotted in Fig.~\ref{Fig3}(d) and find almost the same behavior to $\mbox{var}[S_{AF}]/L_x$. 
From $S_{AH}$ plotted in Fig.~\ref{Fig3}(e), we find that for small $p$, $S_{AH}\propto L_x$, and interestingly enough for large $p$, $S_{AH}$ approaches unity, $S_{AH}=1$. 
It indicates two-fold degenerate states on the upper rough boundary. 
We initially include $S^u_{x}$ in the set of stabilizer generators.
We perform measurement only on  bulk degrees of freedom, and therefore $S^u_x$ remains as a stabilizer generator in the whole process. 
Since $S^u_x$ flips all spins on the upper rough boundary, it contributes to $S_{AH}$\cite{Fattal2004}, as $S_{AH}=1$. 
For large $p$, interplay between bulk-boundary entanglement separation and the stabilizer $S^u_x$ on the rough boundary indicates that the rough boundary state is a cat state (GHZ state such as, $\frac{1}{\sqrt{2}}(|011\cdots 0\rangle\pm |100\cdots 1\rangle$)), where $S^u_x$ (parity) operator stabilizes the state with eigenvalue $1$ or $-1$, i.e., implying a superposition of the SSB states on the boundary. 
This phenomenon is similar to the two-fold degenerate $Z_2$ SSB state in the well-studied transverse field Ising chain.
The SSB indicated by $S_{AH}\to 1$ is a signal of  emergence of the Higgs=SPT phase. 

\begin{figure}[t]
\begin{center} 
\includegraphics[width=8cm]{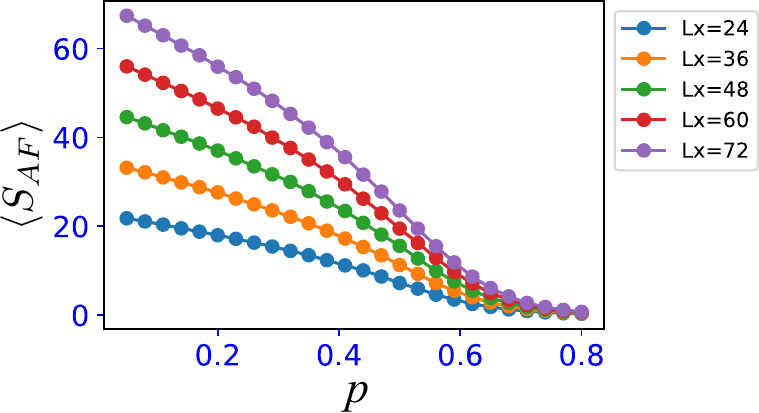}  
\end{center} 
\caption{System size dependence of $S_{AF}$, where we averaged over all measurement samples.
We used $3\times 10^3$ samples. We fix $L_y=6$.}
\label{FigA4}
\end{figure}
Finally, We show the behaviors of the average of entanglement entropy $S_{AF}$ over the measurement samples for different system sizes. 
The result is plotted in Fig.~\ref{FigA4}. 
For small $p$, $S_{AF}\propto L_x$ similarly to the case of $S_{AH}$. 
For large $p$, $S_{AF}$ approaches zero, indicating the strict entangle separation between the bulk and the rough boundary. 


\section{Search for a bulk transition}
Final issue is to search a bulk transition and clarify relation between the bulk and boundary transitions. 
To this end, to investigate the size of $\tilde{A}_v$ clusters $(\tilde{A}_{v_1}\tilde{A}_{v_2}\cdots)$ in the set of stabilizer generators 
is quite useful.
The Gauss law is destroyed at sites adjacent to $\sigma^z$-measurement, whereas the stabilizer of each cluster comprises $\sigma^x_\ell$'s on its boundary, and therefore every $\tilde{A}_v$ cluster measures its total inside charge.
In the critical regime of the transition from the deconfined (toric code) to Higgs phases, we expect the scale invariance, and emergence of various sizes of clusters.
Simple consideration reveals (frequency of clusters) $\propto$ (size of cluster)$^{-1}$.
Probability distribution of size of clusters can be measured by the number of $\sigma^x_\ell$ in the corresponding stabilizer generator.

\begin{figure}[t]
\begin{center} 
\includegraphics[width=7cm]{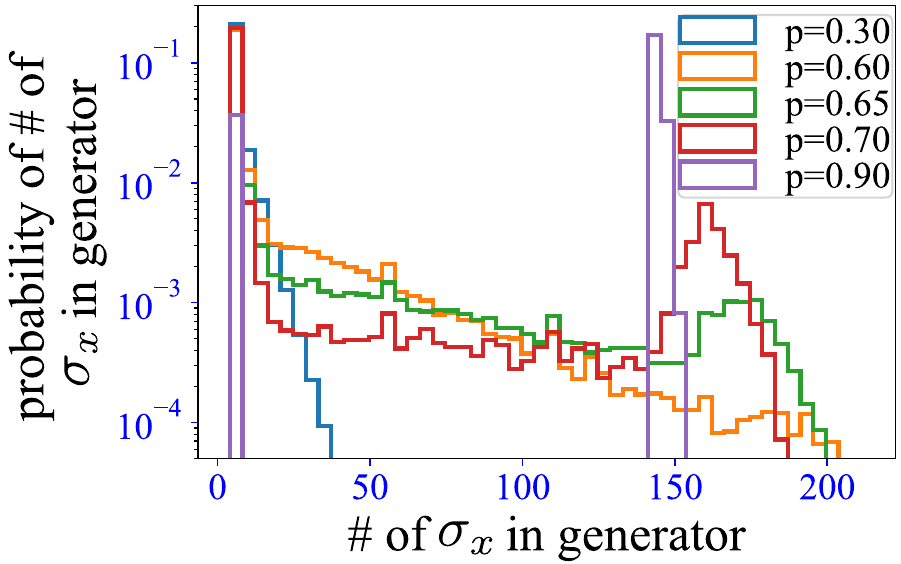}  
\end{center} 
\caption{Histogram of the number of $\sigma_x$ in each generator of stabilizers for $0.3\le p \le 0.9$. We omit the data point for zero number of $\sigma^x$ and $S^u_x$. We used
$3\times 10^3$ samples.}
\label{Fig4}
\end{figure}

The numerical results are displayed in Fig.~\ref{Fig4}.
For small $p=0.30$, the size of clusters is rather small, and for large $p=0.9$, 
all clusters have almost the same magnitude.
For that regime, we expect that large rectangular-shaped clusters tend to emerge.
In between, various size of clusters emerge, and the system for $p=0.60$ seems scale-invariant.
Therefore, we expect that the bulk transition from deconfined (toric code) to Higgs phases takes place at $p=p_b\simeq 0.60.$
Topological order of the initial state gets lost at $p=p_b$, as the efficiency of the operator $\prod_{\rm periodic}\sigma^x_\ell$ is destroyed by `voids' inside $A_v$ clusters, where `periodic' denotes a closed loop on the dual lattice in the periodic direction ($x$-direction) of the cylinder. 
In fact very recently, some related problem was studied in \cite{Botzung}, in which stability of logical qubits in toric code against measurement is discussed by referring percolation theory. The critical probability for breakdown of the logical qubit operator is estimated as $p=0.5$ there.
Our estimation of $p_b$ might approach that value for sufficiently large systems.
It is an interesting future problem to study the relation between our observation about bulk phase transition and the breakdown of logical qubit operator. 

The above result implies that the bulk transition observed in the present protocol takes places earlier than the rough boundary phase transition indicated by the peak of var$(\chi_{SG})/L_x$.
We also observed the variance of the squared expectation value of the local $\tilde{A}_v$, the results of which supports the existence of bulk transition, shown in Appendix.

Summarizing the results, we expect that the bulk phase transition takes place first at $p_b$ and then the rough-boundary phase transition at $p_c$, which is naturally regarded as the SPT transition. 
This observation indicates that the regime $p_b<p<p_c$ is neither topological nor SPT.
As a possible physical picture, with increasing $p$, incoherent proliferation of e-anyon (charge particle in the gauge-Higgs model) occurs first inducing the breakdown of the long-range entanglement at $p=p_b$, and then, further increase in $p$ results in the condensation of e-anyon with the boundary long-range order as the SPT order. 
A similar physical picture has been shown in recent studies \cite{Fan2023,Wang2023}.
We obviously cannot deny the possibility that the above two transitions take place simultaneously.\\ 

\section{Conclusion}
This work clarified the bulk-measurement-induced boundary phase transitions in the generalized toric code system on cylinder geometry with the rough boundaries. 
Our numerical results elucidated the emergence of the SG LRO on the rough boundaries. 
The criticality was estimated in detail. 
The SSB with LRO and the degeneracy on the rough boundaries in our toric code system under projective measurement indicate that the Higgs=SPT phase is produced by the local measurement on the bulk starting with the initial toric code (deconfined) state. 

There are various interesting future directions.
One of them is to study a similar system that is closely related to a subsystem code~\cite{Poulin2005,Bacon2006,Wildeboer2022}.

\section*{Acknowledgements}
This work is supported by JSPS KAKENHI: \\
JP23K13026(Y.K.) and JP23KJ0360(T.O.). 

\appendix

\section*{Appendix: Additional numerical results}
We show additional numerical results for the measured toric code system with cylinder geometry by using the efficient numerical stabilizer algorithm.

\begin{figure}[t]
\begin{center} 
\includegraphics[width=8.0cm]{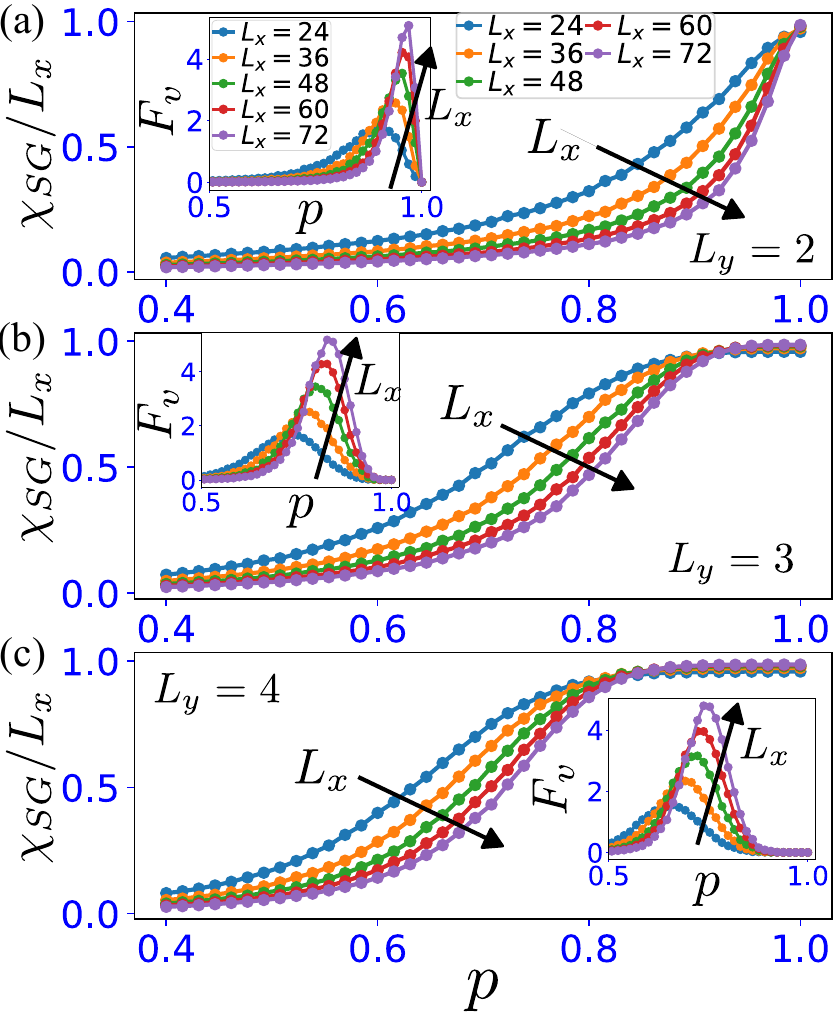}  
\end{center} 
\caption{ 
$L_y$ dependence of spin glass order $\chi_{SG}/L_x$: (a) $L_y=2$, (a) $L_y=3$, and (c) $L_y=4$.
Each inset panel represents $F_v$.
We used $4\times 10^3$ samples.
}
\label{FigA3}
\end{figure}

\begin{figure}[t]
\begin{center} 
\includegraphics[width=8.8cm]{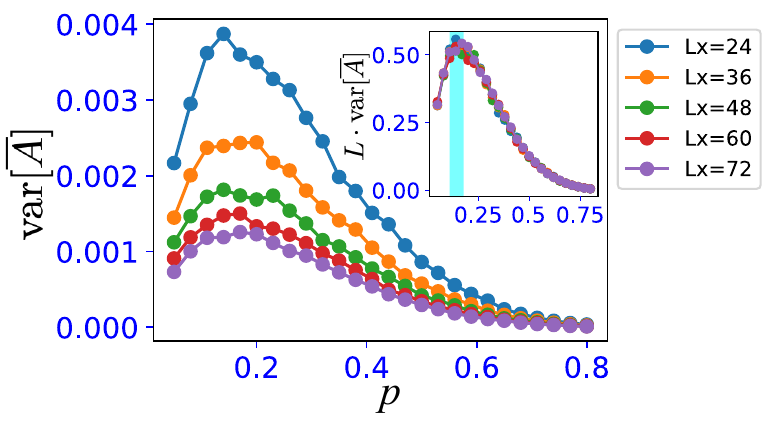}  
\end{center} 
\caption{The sample-to-sample variance of ${\bar A}$. The inset displays the $L\cdot{\bar A}$. 
We used $3\times 10^3$ samples. We fix $L_y=6$.}
\label{Fig10}
\end{figure}
\subsection{$L_y$-dependence of rough-boundary phase transition}
We observe effects of the system size in the $y$-direction, $L_y$, to the rough-boundary phase transition. 
The $L_y$-dependence of $\chi_{SG}$ are displayed in Figs.~\ref{FigA3}(a)-\ref{FigA3}(c) for $L_y=2,3$ and $4$. 
For all values of $L_y$,  the value of $\chi_{SG}/L_x$ increases more rapidly as a function of  $p$ for larger $L_x$. 
Furthermore, for $L_y=2$ [Fig.~\ref{FigA3}(a)], we observe that the data of the various $L_x$'s do not cross with each other except  $p=1$.
This indicates non-existence of  boundary phase transition for a finite $p$, while for larger $L_y$'s, the crossing point shifts to the lower value of $p$ as shown in Figs.~\ref{FigA3}(b) and \ref{FigA3}(c). 
Also, the insets of each panels in Figs.~\ref{FigA3}(a)-\ref{FigA3}(c) show the behavior of $F_v$. 
These results indicate the existence of the continuous phase transition on the rough boundary. 
However for the $L_y=2$ case [Fig.~\ref{FigA3}(a)], the peak of $F_v$ approaches  $p=1$, indicating that the transition point approaches  $p=1$ for the thermodynamics limit.

These numerical results indicate that the SG-ordered phase itself disappears as decreasing $L_y$, corresponding to (1+1)-D system. 
The reason is that in the Hamiltonian formalism, (1+1)-D gauge-Higgs model with periodic boundary conditions has the unique ground state in the `Higgs regime' \cite{Kogut1979,Verresen2022}, and the model is nothing but a specific limit of the cluster spin model with periodic boundary conditions (i.e., no boundaries in the $x$-direction.)
Our numerical result is consistent with the consideration by the Hamiltonian formalism.


\subsection{Variance of squared $\tilde{A}_v$}
We further present numerical study about the expectation value of squared star operator in our protocol, 
$
{\bar A}=\frac{1}{N_v}\sum_{v}\langle \psi|\tilde{A}_{v}|\psi\rangle^2
$, where $N_v=L_x(L_y-1)$.
In our protocol, the local $\sigma^z$-measurement removes the initial $\tilde{A}_v$'s, and thus, its number monotonically decreases for larger $p$. 
In the toric code, this means the lack of the Gauss law and discontinuity of electric flux.
Then, we  calculate the sample-to-sample variance of it, $\mbox{var}[{\bar{A}}]$. 
Its system-size and $p$-dependence are plotted in Fig.~\ref{Fig10}. 

The moderate peaks emerge and their height is smaller for larger $L_x$. 
The reason for it is simple:
Since the variance is calculated from the value of $\langle \psi|\tilde{A}_v|\psi\rangle^2$ averaged  for all vertices $v$ in each sample, the variance of the mean is inversely proportional to $L_x$ although these values are much small.
In fact as shown in the inset of Fig.~\ref{Fig10},  $\mbox{var}[{\bar{A}}]\times L_x$ exhibits no system-size dependence. 
This behavior can support the presence of bulk phase transition as explained by observing the scale-free distribution of the size of the product of $\tilde{A}_v$ shown in Fig.~\ref{Fig4} in the main text. The value of $p$ on the moderate peak of $\mbox{var}[{\bar{A}}]\times L_x$ is estimated as $p\sim 0.14$. 
This value is small compared to $p_b$ and $p_c$. This result implies that after weak fluctuation of ${\bar A}$, the scale free phenomena for the size of stabilizer constituted by the product of $\tilde{A}_v$ (e-anyon cluster proliferation) emerges.


\begin{thebibliography}{99}
\bibitem{LRB}
Even if we select $\displaystyle{S^{l}_x= \prod_{\ell\in {\rm LRB}}\sigma^x_{\ell}}$ (where LRB stands for the lower rough boundary) instead of $S^{u}_x$, the stabilized physical state does not change because $S^{l}_x$ is not independent to $S^{u}_x$ since $S^{l}_x$ is connected to $S^{u}_x$ by multiplying all $A_v$ initial stabilizer generators.
\end{thebibliography}
\end{document}